\documentstyle[psfig,12pt]{article}

\newcommand{\be}{\begin{equation}}
\newcommand{\ee}{\end{equation}}
\newcommand{\ba}{\begin{eqnarray}}
\newcommand{\ea}{\end{eqnarray}}

\begin{document}

\begin{center}
{\Large Search for Direct Stress Correlation Signatures \\of the Critical Earthquake Model}
\end{center}
\vskip 0.5cm
\begin{center}
G. Ouillon$^1$ and D. Sornette$^{1,2}$
\end{center}
\vskip 0.5cm

\noindent $^1$ Laboratoire de Physique de la Mati\`{e}re Condens\'{e}e,
CNRS UMR 6622 and Universit\'{e} de Nice-Sophia Antipolis, Parc Valrose,
06108 Nice, France\\
$^2$ Department of Earth and Space Sciences and Institute of
Geophysics and Planetary Physics, University of California, Los Angeles,
California \\(e-mails: ouillon@aol.com and sornette@moho.ess.ucla.edu)

\vskip 2cm
\begin{abstract}

We propose a new test of the critical earthquake model based on the 
hypothesis that precursory earthquakes are ``actors'' that create
fluctuations in the stress field which exhibit an increasing correlation length
as the critical large event becomes imminent. 
Our approach constitutes an attempt to build a more physically-based
cumulative function in the spirit of but improving on the cumulative Benioff strain used in 
previous works documenting the phenomenon of accelerated seismicity.
Using a space and time
dependent visco-elastic Green function in a two-layer model of the Earth
lithosphere, we compute the spatio-temporal stress fluctuations 
induced by every earthquake precursor and estimate, 
through an appropriate wavelet transform, the contribution of each event to the correlation
properties of the stress field around the location of the main shock at different
scales. Our physically-based definition of the cumulative stress function 
adding up the contribution of stress loads by all earthquakes preceding a main shock
seems to be unable to reproduce an acceleration of the cumulative
stress nor an increase of the stress correlation length similar to those observed previously
for the cumulative Benioff strain. Either earthquakes are ``witnesses'' of large
scale tectonic organization and/or the 
triggering Green function requires much more than just visco-elastic stress transfers.

\end{abstract}

\pagebreak
\section{Introduction}

Numerous reports of 
precursory geophysical anomalies preceding earthquakes have fueled the hope for
the development of forecasting or predicting tools. 
The suggested anomalies take many different forms and relate to many 
different disciplines such as seismic wave propagation, 
chemistry, hydrology, electro-magnetism and so on. The
most straightforward approach consists in using patterns of
seismicity rates to attempt to forecast future large events 
(see for instance [Keilis-Borok and Soloviev, 2002] and references therein).

Spatio-temporal patterns of seismicity, such as anomalous bursts of aftershocks, 
quiescence or accelerated seismicity, are
thought to betray a state of progressive damage or of organization
within the earth crust preparing the stage for a large earthquake.
There is a large literature reporting that large events have been
preceded by anomalous trends of seismic activity both in time
and space. Some works report that seismic activity increases as an inverse power
of the time to the main event (sometimes refered to as an inverse Omori law for relatively
short time spans), while
others document a quiescence, or even contest the existence of such
anomalies at all. 

There is an almost general consensus that
those anomalous patterns, if any, are likely to occur within days to weeks before
the mainshock and probably not at larger time scales [Jones and Molnar, 1979]. 
With respect to spatial structures, the precursory patterns
are very often sought or observed in the immediate vicinity of the
mainshock, i.e., within distances of a few rupture lengths from the
epicenter. The most famous observed pattern is the so-called
doughnut pattern. Thus, in any case, both temporal and spatial
precursory patterns are usually thought to take place at short distances
from the upcoming large event. 

In the last decade, a different concept has progressively emerged according
to which precursory seismic patterns may occur up to decades preceding large 
earthquakes and at spatial distances many times the main shock
rupture length. This concept is rooted in the theory of critical phenomena
(see [Sornette, 2000] for an introduction and a review adapted
to a general geophysical readership) and has been documented and advocated
forcefully by the russian school [Keilis-Borok, 1990; Keilis-Borok and Soloviev, 2002].
Probably the first report by Keilis-Borok and Malinovskaya [1964]
of an earthquake precursor (the premonitory increase
in the total area of the ruptures in the earthquake sources in a medium magnitude
range) already featured very long-range correlations (over 10 seismic source lengths)
and worldwide similarity.
More recently, Knopoff et al. [1996] have also discovered a surprising long-range
spatial dependence in the increase of medium range magnitude seismicity prior to 
large earthquakes in California.
From a theoretical point of view, its seismological roots dates back
to the branching model of [Vere-Jones, 1977]. A few years later,
All\`egre et al. [1982] proposed a percolation model of
damage/rupture prior to an earthquake, emphasizing the multi-scale
nature of rupture prior to a critical percolation point. Their model is actually nothing
but a rephrasing of the real-space
renormalization group approach to a percolation model 
performed by Reynolds et al. [1977]. Similar ideas were also explored
in a hierarchical model of rupture by Smalley et al. [1985]. 
Sornette and Sornette [1990] proposed an observable consequence of the
critical point model of All\`egre et al. [1982] with the goal of verifying the
proposed scaling rules of rupture. Almost simultaneously but following
apparently an independent line of thought, 
Voight [1988; 1989] introduced
the idea of a time-to-failure analysis in the form of an empirical second order
nonlinear differential equation, which for certain values of the
parameters leads to a time-to-failure
power law of the form of an inverse Omori law. This was used and tested later for 
predicting volcanic eruptions. Then, 
Sykes and Jaum\'e [1990] performed the first empirical study reporting and
quantifying with a specific law an acceleration of seismicity prior to
large earthquakes. They used an exponential law to describe
the acceleration and did not use or discuss the concept of a critical
earthquake. Bufe and Varnes [1993] re-introduced a time-to-failure
power law to model the observed accelerated seismicity quantified by the so-called
cumulative Benioff strain. Their justification of the power law was a mechanical
model of material damage. They did not refer nor discussed the
concept of a critical earthquake. Sornette and Sammis [1995]
was the first work which reinterpreted the work of Bufe and Varnes [1993] 
and all the previous ones reporting accelerated seismicity within the model
of a large earthquake viewed as a critical point in the sense of  
the statistical physics framework of critical phase transitions. 
The work of Sornette and Sammis [1995] generalized significantly 
[All\`egre et al., 1982; Smalley et al., 1985] in that their proposed
critical point theory does not rely on an irreversible damage but refers
to a more general self-organization of the stress field prior to large
earthquakes. In addition, using the insight of critical points in
rupture phenomena, Sornette and Sammis [1995] proposed to enrich the power law
description of accelerated seismicity
by considering complex exponents (i.e., logperiodic corrections to
scaling [Newman et al., 1995; Saleur et al., 1996; Johansen et al., 1996; 2000; Ouillon
and Sornette, 2000]). 
This concept has been elaborated theoretically to accomodate
both the possibility of critical self-organization (SOC) and the critical
earthquake concept [Huang et al., 1998]. 
Bowman et al. [1998] gave empirical flesh to these ideas by showing that 
all large Californian
events with magnitude larger than $6.5$ are systematically preceded by a
power-law acceleration of seismic activity in time during several
decades, in a spatial domain about $10$ to $20$ times larger than the
impending rupture length (i.e., of a few hundreds kilometers). The large
event could thus be seen as a temporal singularity in the seismic
history time-series. Such a theoretical framework
implies that a large event results from the collective behaviour and
accumulation of many previous smaller-sized events. Similar
results were also obtained by Brehm and Braile [1998, 1999] for
other earthquakes. Jaum\'e and Sykes [1999]
have reviewed the critical point concept for large earthquakes 
and the data supporting it. The additional results of
Ouillon and Sornette [2000] on mining-induced seismicity, and
Johansen and Sornette [2000] in laboratory experiments, brought
similar conclusions on systems of very different scales, in good
agreement with the scale-invariant phenomenology reminiscent of systems
undergoing a second-order critical phase transition. In this picture, the
system is subjected to an increasing external mechanical sollicitation.
As the external stress increases, micro-ruptures occur within the
medium which locally redistribute stress, creating stress fluctuations
within the system. As damage accumulates, fluctuations interfere and
become more and more spatially and temporally correlated, i.e.,
there are more and more, larger and larger domains that are
significantly stressed, and thus larger and larger events can occur at
smaller and smaller time intervals. This accelerating spatial smoothing
of the stress field fluctuations eventually culminates in a rupture
which size is of the order of the size of the system. This is the final
rupture of laboratory samples, or earthquakes breaking through the
entire seismo-tectonic domain to which they belong. This concept was
verified in numerical experiments led by Mora et al. [2000, 2001], who showed that the
correlation length of the stress field fluctuations increases
significantly before a large shock occurred in a discrete numerical model. 
More recently, Bowman
and King [2001] have shown with empirical data that, in a large domain including
the impending major event, similar
to the critical domain proposed in Bowman et al. [1998], the maximum
size of natural earthquakes increased with time up to the main shock. If
one assumes that the maximum rupture length at a given time is given by
(or related to) the stress field correlation length, then this last work
shows that this correlation length increases before a large rupture.
Sammis and Sornette [2002] summarized the 
most important mechanisms creating the positive feedback at the possible
origin of the power law acceleration. They also 
introduced and solved analytically a novel simple model based on [Bowman and King, 2001]
of geometrical positive feedback in which the stress shadow cast by the
last large earthquake is progressively fragmented by the increasing
tectonic stress. Keilis-Borok [1990] has also used repeatedly the concept of 
a ``critical'' point, but in a broader and looser sense than the restricted
meaning of the statistical physics of phase transitions
(see also [Keilis-Borok and Soloviev, 2002]
for a review of some of the russian research in this area).
The situation is however more complicated when the strain (rather than the stress)
rate is imposed; in that case, the system may not evolve towards a critical point.
The unifying view point is to ask whether the dissipation of energy by the
deteriorating system slows down or accelerates. The answer
to that question depends on a competition
between the nature of the external loading, the evolution of the deterioration
within the system and how the resulting evolving mechanical characteristics
of the system feedback on the external loading conditions. For a constant
applied stress rate, the dissipated energy rate diverges in general in finite
time leading to a critical behavior. For a constant strain rate, the answer
depends on the damage law [Sornette, 1989a]. For a constant applied load,
Guarino et al. [2002] find a critical behavior of the 
cumulative acoustic energy both for wood and fiberglass, with an exponent $\approx -0.26$
which does not depend on the imposed stress and is the same as for 
a constant stress rate.

For the Earth crust, the situation is in between
the ideal constant strain and constant stress
loading states and the critical point may emerge
as a mode of localization of a global input of energy to the system.
The critical point approach leads to an alternative physical picture of the
so-called seismic cycle. From the beginning of
the cycle, small earthquakes accumulate and modify the stress field
within the Earth crust, making it correlated
over larger and larger scales. When this correlation length reaches the
size of the local seismo-tectonic domain,
a very large rupture may occur, which, together with its early aftershocks,
destroys correlations at all spatial scales.
This is the end of the seismic cycle, and the beginning of a new one,
leading to the next large event.
As earthquakes are distributed in size according to the
Gutenberg-Richter law, small to medium-sized events are
negligible in the energetic balance of the tectonic system, which
is dominated by the largest final event. However,
they are ``seismo-active'' (actors) in the sense that their occurrence prepares
that of the largest one.
The opposite view of the seismic cycle is to consider that it is the
large scale tectonic plate displacements which
dominates the preparation of the largest events, which can be modelled
to first order as a simple stick-slip
phenomenon. In that case, all smaller-sized events would be
seismo-passive (witnesses) in the sense that they would reflect
only the boundary loading conditions acting on isolated faults without
much correlations from one event to the other.

Notwithstanding these works,
the critical earthquake concept remains a working hypothesis 
[Gross and Rundle, 1998]: 
from an empirical point of view, the reported analyses
possess deficiencies and a full statistical analysis establishing the confidence level
of this hypothesis remains to be performed. 
In this vain, Zoller et al. [2001] and Zoller and Hainzl [2001, 2002] have
recently performed novel and systematic spatiotemporal 
tests of the critical point hypothesis for
large earthquakes based on the quantification of the predictive power of both
the predicted accelerating moment release and the growth of the spatial
correlation length. These works give fresh support to the concept.

In order to prove (or refute) that a boundary between tectonic plates is
really a critical system, the use of proxies to check the
existence or absence of a build-up of cooperativity
preparing a large event in terms of cumulative (Benioff) strain should ideally
be replaced by a direct measure of the stress field. Indeed, 
one should measure
the evolution of the stress field in space and time in such a region,
compute its spatial correlation function, deduce
the spatial correlation length, and show that it increases with time as
a power-law which defines a singularity when the
mainshock occurs. Unfortunately, such a procedure is far beyond our
technical observational abilities. 
First, it is well-accepted that large earthquakes nucleate at a depth of
about $10-15km$, so it is likely that stress field values and
correlations would have to be measured at such a depth to get
an unambiguous signature. Moreover, the
tensorial stress field would have
to be measured with a high resolution in order to show evidence of
a clear increase of the correlation length.
As those measurements are clearly out of reach at present, we propose here a
simplified method to approach such a goal.
We will then consider the 4 last largest recent events that have
occurred in Southern California (Loma Prieta (1989), Landers (1992),
Northridge(1994), Hector Mine(1999)), and test if such a critical
scenario is likely to have taken place prior to their occurrence.

Our approach constitutes an attempt to build a more physically- or mechanically-based
cumulative function in the spirit of the cumulative Benioff strain used in 
previous works documenting the phenomenon of accelerated seismicity.

\section{General methodology}

As direct stress measurements of sufficient extent for the purpose
of estimating a correlation legnth are clearly out of reach, our
goal is to estimate indirectly
the stress distribution and its evolution with time within the crust through
a numerical procedure based on instrumental seismicity.

Estimating the spatial stress history within a tectonic
domain requires three different kinds of data: the first one consists in
the knowledge of the far-field stress and/or strain
imposed on the system. The second one consists in the accurate knowledge
of the Earth's crust structure and rheology.
The third one consists in the knowledge of the sources of internal
stress fluctuations, which are mainly related to
earthquake occurrence, whatever their size. The time evolution of the
spatial structure of the stress field is thus created by
the superposition of both far-field and internal contributions,
coupled with the rheological response of the system
(which can be quite complex). Despite its apparent simplicity, the first
kind of data is still largely under debate.
For example, very different scenarios are still proposed for the tectonic
loading of the San Andreas fault system.
Moreover, the determination of the precise boundaries of the system
remain a subjet of controversy and research due to the
complexity associated with the fractal hierarchical organization
of tectonic blocks [Sornette and Pisarenko, 2002]. Fortunately, 
the critical point theory ensures that one needs only to consider
the correlation function of internal fluctuations, which are the ones
related to earthquakes occurrence, and not the large scale effects of
the boundary conditions as long as they vary slowly on the time scale
of the seismic cycle. This is why we
will not further consider boundary conditions anymore here. 

We shall thus use earthquake catalogs as the source of information
available to qualify and quantify stress field
fluctuations. Usual catalogs contain
parameters such as earthquake location
(longitude, latitude, depth), origin time and magnitude. For example,
the CALTECH catalog that we use here is considered to be
complete since 1932 for magnitudes larger than about 3.5. Unfortunately,
these informations are not sufficient for quantifying the spatial stress
perturbations due to a given seismic event. Two major ingredients are lacking.
First, we must
know the details of the rupture mechanism. This includes size (length
and width), strike and dip of the fault plane,
as well as the slip distribution upon it (in amplitude and direction).
Those informations are usually only available for
spatially and temporally restricted catalogs (but which can cover a
large magnitude interval), or for more extended
catalogs but only for shocks of large magnitudes (for example the
Harvard catalog for shocks of magnitude larger
than 5.5). As there are so few such events diluted in a very large
spatial and temporal domain, it is clear that we will get in this way information
on the stress field structure only at very large scales. If we consider
all events in a catalog, we should be able to get insight into smaller
scales (as events are much more numerous and have shorter rupture
lengths), but would lack the information on the source parameters. 
We shall opt for the
option using all the observed and complete seismicity, 
and will define in the next section a simplified Green function
giving the spatial structure of the internal stress fluctuations due to
an event of any size occurring anywhere at any time within our system. A
drastic consequence will be that this Green function will be a scalar
rather than the correct tensorial structure which would be 
accessible if we knew the details of the
rupture. Our hope is that if the critical nature of rupture is a strong
property, it should be detectable even with such an approximation.
In addition, the superposition effect of scalars gives in general stronger fluctuations
than for higher dimensional objects such as moment tensors
due to the lack of dispersion along 
several possible directions. The existence, if any, of an increasing correlation
of the stress field should thus be detectable more easily, even if not exact quantitatively.

In order to estimate reliably the stress fluctuations and their
evolution with time, we also need an accurate rheological model of the
local lithosphere, including knowledge of elastic constants and
relaxation times for the viscous layers. These latter ingredients can be
deduced from geophysical investigations, at least on a large scale. Of
course, the more accurate will be this model, the more difficult and
lengthy will be the estimations of the stress field perturbations, which
would necessitate the use of a finite elements or boundary elements
codes. As the rheological behavior of the Earth crust and lithosphere's
material can be quite complex, we shall use in the following a
simplifed rheological model which captures the essential features of
stress transmission and relaxation within a viscoelastic layered medium.

The methodology used in this work is the following: we first choose a
recent large event (to ensure a sufficiently large catalog of possible
precursor events, both in time and number), occuring at time $T_0$ and
location $P_0$. We read every event in the catalog which preceeds it,
and compute the spatio-temporal stress fluctuations it induces in the
whole space. We also estimate, through an appropriate wavelet transform
(see below), the contribution of each event to the correlation
properties of the stress field around location $P_0$ at different
scales. This will provide us with the correlation length of the stress field
around $P_0$ and its evolution with time, up to the time of occurrence of the
large event.

\section{Construction of the Green function}

We will first consider the stress field due to a seismic source in a 3-D
elastic, infinite and isotropic medium. As catalogs do not provide us with
all the parameters needed to compute accurately the exact elastic solution, we will
make the following assumptions.
\begin{itemize}
\item[(i)] We will consider that each source is isotropic and that the stress
perturbation is positive with radial symmetry around the source.
\item[(ii)] This stress perturbation $\sigma_L(r)$ is assumed to decay from the source as:
\begin{equation}
\sigma_L(r) = {(L/2)^3 \over (L/2)^3+r^3}~,
\label{mgksl}
\end{equation}
where $L$ is the linear size of the source (which plays the role of the
rupture length in real events), and $r$ is the distance from the
source. 
\end{itemize}
The size $L$ is determined empirically using a statistical
relationship between magnitudes and rupture lengths established for strike slip
faults in California [Wells and Coppersmith, 1994]:
\begin{equation}
\log(L) = -2.57 + 0.62 \times M_l~,
\label{ngbnjbklld}
\end{equation}
where $M_l$ is the local magnitude and $L$ is expressed in kilometers. To ensure
that all earthquakes are treated on the same footing, this
statistical relationship is also used for the events for which the information
on the rupture plane is available. Note that the computed stress $\sigma_L(r)$ is
always positive and does not depend on azimut, so that it does not really
define a genuine stress, but can be interpreted as a kind of influence function, with $L$
playing the role of the size of the area in which a shock will possibly
influence following events.

We now take into account that the source does not occur in a purely homogeneous
elastic medium, but in a two-layers viscoelastic one. The upper layer is
considered as a viscoelastic medium with relaxation time $\tau _1$. 
The lower layer is also taken as a viscoelastic medium (possibly
semi-infinite) with relaxation time $\tau_2 < \tau_1$. We assume that 
earthquakes are localized within the upper (more brittle) layer, and that the
quantity of interest is the scalar stress field measured
in this layer, taken constant in the vertical dimension so as to ensure 
that the stress field is two-dimensional within the horizontal plane.
The thicknesses of the layers and the existence of free surfaces are
embodied in phenomenological constants defined
below. The depths of the events is taken identical and we neglect any vertical 
variation. This amounts to calculate the stress field at this nucleation depth.

The rupture and relaxation of the stress field in the two-layer system is 
modeled as follows. Once an event occurs in
the upper layer, the instantaneous elastic solution for the stress field
is given by expression (\ref{mgksl}). Then, both layers begin to flow by
viscous relaxation. The lower layer flows faster, due to a smaller
relaxation time associated with a more ductile rheology. The effect of
this viscous relaxation is to progressively load the upper layer and
thus creates a kind of post-seismic rebound. 
This loading effect computed in the upper layer
is assumed to be described by a function of the type:
\begin{equation}
f(r,t) = \sigma_L(r) [1-C\exp (-t/\tau_2)]~H(t)~,
\label{mgmslls}
\end{equation}
where $\sigma_L(r) $ is the elastic isotropic solution given by
(\ref{mgksl}), $C$ is a constant which
quantifies the maximum quantity of stress which is transfered in the
upper layer, and which depends on the geometry of the problem. If $C=0$, no
transfer occurs. $H(t)$ is the Heavyside function which ensures that the
stress fluctuation becomes non-zero once the event has occurred. Here,
$t$ is the time elapsed since the seismic event. At the same time, the stress
also relaxes in the upper layer, at a rate which varies as
$\exp(-t/\tau_1)$. This relaxation takes into account the usual viscous processes
as well as the effect of micro-earthquakes which dissipate mechanical
energy. 

As both relaxations occur simultaneously, the evolution of the stress field in
the upper layer is given by the sum of two contributions: (1) the 
direct relaxation $\sigma_L(r) \exp(-t/\tau_1)$ of the instantaneous elastic
stress load in the upper layer due to the event and (2) the
convolution of the time-derivative of $f(r,t)$ with the exponential relaxation function
$\exp(-t/\tau_1)$ in the upper layer. This second contribution sums over
all incremental stress sources $df(r,t)/dt$ per unit time
in the upper layer stemming from the relaxation
of the lower layer. After some algebra, we get
the stress perturbation induced by an earthquake under the form
\begin{equation}
\sigma(r,t) = \frac{(L/2)^3}{(L/2)^3+r^3} \left[\exp (-t/\tau_1)
+B\frac{\tau_1}{\tau_1-\tau_2}\left(\exp (-t/\tau_1)-\exp (-t/\tau_2)\right)\right]~,
\label{mgjslala}
\end{equation}
where $r$ and $t$ are respectively the horizontal distance from the
source and the time since the occurrence of the earthquake.
The constant $B$ represents the relative contribution to the stress
field in the upper layer due to the delayed loading
by the slow viscous relaxation of the lower layer that has
been loaded by the instantaneous elastic stress
transfer at the time of the earthquake compared with the direct relaxation of the
elastic stress created directly in the upper layer. The numerical value of $B$
is difficult to ascertain as it depends strongly on the geometry of the 
layers as well as on their rheological constrast. We expect both contributions
to be of the same order of magnitudes and, in the following, we shall take $B=1$.

The Green
function defined here is a rough approximation of what
really takes place within the crust and the lithosphere, but it nonetheless
captures qualitatively the overall evolution of the stress field. One could
raise the criticism that it does not feature any azimutal dependence of the stress field
perturbation but, as we already stated, this is done in view of the absence
of detailed information on the source mechanisms of the events.
On the other hand, as stated above, the use of an isotropic stress field is expected
to lead to an overestimation of the
correlation length, that is, to an amplification of the signal we are searching for.
While we cannot provide a rigorous proof of this statement, it is based on the analogy
between percolation and Anderson localization 
[Souillard, 1987, Sornette, 1989b,c]: the first phenomenon describes the transition
of a system from conducting to isolating by the effect of the
addition of positive-only contributions;
The second phenomenon refers to the transition from conducting to isolating when
taking into account the ``interferences'' between the positive, negative 
and more generally phase-dependent amplitudes of the electronic quantum wave functions.
In this later case, the transition still exists but is much harder to obtain
and to observe. In the future, it may nevertheless
be interesting to check this point and test a
generalization of the present model in which a random source orientation is 
chosen for each event and the angular dependence of the associated double-couple 
stress is taken into account. 

The Green function we propose also assumes a complete decoupling between
space and time, so that viscous relaxation does not exhibit any diffusive
pattern. Indeed, such a diffusion would imply an increase of the size of
the influence area with time. As the amplitude of the stress
signal decreases exponentially with time, we believe that this mechanism is not
crucial (because too slow and too weak
in amplitude) in order to obtain and measure an increase of the stress field
correlation length, if any. Another assumption of our rheological model is that
the viscoelastic component is linear, allowing to clearly define
relaxation times. This ingredient allows us to define a
simple and convenient computation procedure to estimate a correlation length,
as discussed in the next section.

The simplified Green function $\sigma(r,t)$ given by (\ref{mgjslala})
has several interesting properties catching the overall physics of the
stress evolution in the upper layer after an event. The elastic prefactor
$\sigma_L(r)$ given by (\ref{mgksl})
implies that the stress perturbation is initially of
order unity within a circle of radius $L/2$, and sharply decreases as
$r^{-3}$ outside this circle. Note that the maximum amplitude
of the stress perturbation
is independent of the size $L$, as the stress drop is thought
to be constant, whatever the size of an event. At any point in the upper
layer, the stress will first be given by the elastic solution. As
$\tau_1 > \tau_2$, the stress at any point in the upper
layer will first increase due to the relaxation of the lower layer, reach a
maximum, and then decrease with time as the upper layer is relaxing too,
but with a longer relaxation time. 

Figure 1 shows such a scenario with
$\tau_1 =10$ years and $\tau_2 =1$ year. The maximum amplitude depends on
the distance between the event and the point where this stress is
measured (as well as on $B$). 

If we now superimpose the contributions of all
successive earthquakes in a catalog, the stress history at any given
point will be a succession and/or superposition of such fast growing and
slowly decaying stress pulses. We thus construct
the cumulative stress function $\Sigma(t)$ defined as
\be
\Sigma(t) = \sum_{i} \sigma(r_i,t_i)~,
\label{mbmjaka}
\ee
where $\sigma(r_i,t_i)$ is given by (\ref{mgjslala})
and $r_i$ and $t_i$ are the distance and the time of event $i$ to the main shock.
For example, Figure 2a shows the stress
history measured at the location of the Landers epicenter due to the
succession of all previous events in the catalog, assuming $\tau_1 =1$
and $\tau_2 =6$ months. Figure 2b shows the same computation
for $\tau_1 = 10$ years, while Figure 2c assumes $\tau_1 = 100$
years. Increasing $\tau_1$ widens the stress pulses, which lead them to
overlap and produces a more continuous stress history. 

The constructions of $\Sigma(t)$ shown in Figure 2a-c are analogous to the cumulative
Benioff strain studied in 
[Bufe and Varnes, 1993; Sornette and Sammis, 1995; Bowman et al., 1998;
Brehm and Braile, 1998; 1999; Jaum\'e and Sykes, 1999; Ouillon and Sornette, 2000],
and are an attempt to improve upon them as we now explain.
They are analogous because they can be seen as similar to the sums of the type
\be
M_q(t) = \sum_{i | t_i<t}~ [M_0(i)]^q~,
\label{mgmlslalq}
\ee
where $M_q(t)$ is a moment generating function of order $q$, $t_i$ and $M_0(i)$
are the time and seismic moments of the $i$-th earthquake and $q$ is an exponent
usually taken between $0$ and $1$. The cumulative Benioff strain is obtained 
as $M_{q=1/2}(t)$ where the sum is performed over all events above a magnitude cut-off
in a pre-defined spatial domain. Taking $q=1$ corresponds to summing the seismic
moments, while taking $q=0$ amounts to simply constructing the cumulative number 
of earthquakes. The constructions shown in Figure 2a-c can be seen as equivalent
to $M_{q=0}(t)$ when
the two following limits hold: (1) all earthquakes in the catalog are so close to
each other that they are all within a distance less than their rupture length
from the point where the stress is calculated (in this case, the elastic
stress perturbation brought by each event is equal to the constant stress drop);
(2) the time difference between
the occurrence of each event and the main shock is significantly less than $\tau_2$ 
such that the time-dependence in (\ref{mgjslala}) can be neglected.

A significant advantage in our
construction of the cumulative stress function $\Sigma(t)$ defined by (\ref{mbmjaka})
compared with the cumulative Benioff strain
resides in the fact that we do not need to specify in advance a spatial domain, a delicate
and not-fully resolved issue in the construction of cumulative Benioff strain functions. 
The definition of the relevant spatial domain is automatically taken into account by
the spatial dependence of the Green function.

Two ingredients are going to modify the observed acceleration of the
Benioff strain when studying the cumulative stress function $\Sigma(t)$ defined 
by (\ref{mbmjaka}).
The first one is that each event contributes a maximum stress perturbation 
equal to the stress drop. In contrast, large events contribute significantly more in
the cumulative Benioff strain as the square-root of their seismic moment and
independently of their distance. There is however a size effect in our calculation of  
$\Sigma(t)$ that reveals itself
at large distances $r_i \gg L_i$, stemming from the magnitude dependence
of the range $L_i$ of the stress perturbation.
According to (\ref{ngbnjbklld}) and using the standard relationship 
between magnitude $M_L$ and seismic moment $M_0$,
$M_L = (2/3) [\log M_0  - 9]$, we obtain $L_i \sim [M_0(i)]^{0.4}$ and thus
$\sigma(r_i,t_i) \sim L_i^3 \sim [M_0(i)]^{1.2}$ for $r_i \gg L_i$. This size effect
has however an almost negligible contribution in generating an acceleration because the stress
field becomes small at large distances. The second ingredient limiting the acceleration 
of the cumulative stress function $\Sigma(t)$ defined by (\ref{mbmjaka}) is the 
relaxation in time which is responsable for the decay observed in Figures 2a-c. The longer
$\tau_1$ is, the smaller is the amplitude of this decay, until $\Sigma(t)$ is replaced
by a staircase in the limit $\tau_1 \to +\infty$. The largest values of $\tau_1$ that
we have explored are significantly larger than the total duration of the catalog
and larger values will not change our results quantitatively.

Another important issue is the contribution of the small events not taken into 
account in the sum (\ref{mbmjaka}). Indeed, the typical area $S(L)$ over which the 
stress redistribution after an event is significant is of the order of the square 
$S(L) \propto L^2$
of the size $L$ of the rupture. If the earthquake seismic moments $M$ are distributed
according to a density Pareto power law $\propto 1/M^{1+\beta}$ with $\beta \approx 2/3$
(which is nothing but the Gutenberg-Richter law for magnitudes translated into moments), 
using the fact that $M \propto L^3$, the density distribution of the areas $S(L)$ is 
also a power law  $\propto 1/S^{1+(3/2)\beta}$ with an exponent $(3/2)\beta \approx 1$.
Thus, the contribution of each class of earthquake magnitudes is an invariant:
small earthquakes contribute as much as large earthquakes to the sum (\ref{mbmjaka}).
Therefore, it seems a priori very dangerous to ignore them in our sum (\ref{mbmjaka}) 
which attempts to detect a build-up of correlation. However, if we assume that the 
physics of self-organization of the crust prior to a critical point is
self-similar, the critical behavior should be observable at all the different
scales and neglecting the contribution of small events should not lead to
a destruction of the signal nor to a modification of 
its relative variations, only to a change in its absolute amplitude.

To sum up, our physically-based definition of the cumulative stress function 
adding up the contribution of stress loads by all earthquakes preceding a main shock
seems to be unable to reproduce an acceleration similar to those observed previously
for the cumulative Benioff strain. This is due to the fact that, conditioned on the
hypothesis of a magnitude-independent stress drop and using standard elasticity, 
the impact of the largest events is not significantly larger than those of smaller events.
In view of this failure, we now attempt another hopefully more robust characterization 
of the critical point model.

\section{Analysis of the structure of the stress field}

Our objective is to determine the
correlation length of the computed stress field in the neighborood of 4
large shocks in California as a function of the time
before their occurrence. In this goal, we are going to analyze
the structure of the stress field around each main shock epicenter to
check whether the
stress fluctuations are increasing or decreasing in size around each main shock
epicenter. In order to extract a robust estimation of the correlation length
of the stress field reconstructed from a limited number of events, we investigate
what spatial scales or wavelengths are
developping around each main shock epicenter, that is, what is
the characteristic scale of the roughness of
the computed stress field. 

An efficient way to achieve such a goal is to perform a $2D$
wavelet transform of the stress field, which acts as a microscope allowing us to focus on
separate scales. As we are interested
only in the spatial structure surrounding the upcoming mainshock (defined
as point $P_0$), we compute wavelet coefficients centered at location $P_0$.
We consider the following wavelet:
\begin{equation}
\frac{1}{a}(2-\frac{r_p^2}{a^2})\exp (-\frac{r_p^2}{2a^2})
\end{equation}
centered at point $P_0$. This ``Mexican hat'' wavelet is the second-order
derivative of the Gaussian function. By construction, it eliminates signals of constant
amplitudes or of constant gradient at scale $a$ or larger. It is thus well
indicated to isolate fluctuations at various chosen scales. $r_p$ is the
distance to point $P_0$, and $a$ is the analyzing scale (the larger $a$,
the larger the width of the wavelet). Note that working with a scale
$a$ means that the corresponding structures have in fact a size $2.2a$
[Ouillon, 1995].

For each time in the stress
field history, the wavelet transform is obtained by convolution of this
function with the
computed spatial stress field, for different values of $a$. If the
resulting wavelet coefficient is close to $0$, this means that the
stress field is uniform or varies linearly around $P_0$, at scale $a$. If
the coefficient is strongly negative, this means that $P_0$ is at or
near a local stress minimum, at scale $a$. If it is strongly positive, this
means that $P_0$ is at or near a stress maximum at scale $a$, indicating that
the stress is both locally high and correlated at that scale. This 
is exactly the property that we want to check. 

Our analyzing procedure is thus the following: we consider
the first event in the catalog. We compute the stress field fluctuation
due to this event at any time and any location through equation (\ref{mgjslala}).
The wavelet transform provides the contribution of this event at any
time to the total wavelet coefficient at any scale $a$ at location $P_0$.
Summing all contributions of successive events (as the rheology we
chosen is linear) up to the major mainshock provides us with the complete
evolution of the scale content of our computed stress field around $P_0$.
From the wavelet coefficient of the cumulative stress field
as a function of scale at a fixed time $t$, we extract the
corresponding correlation length $\rho(t)$ as the scale
corresponding to the maximum coefficient, multiplied by $2.2$. If the
critical point hypothesis is correct, $\rho(t)$ should behave as
\begin{equation}
\rho(t) = A+C(T_0-t)^{-\nu}~,
\end{equation}
where $\nu$ is a positive critical exponent. Note that, due the very
small rupture size $L$ for small earthquakes, and as the
scale $a$ varies
from $1$ to $100 km$, it would be necessary to grid a very large domain
(of few  hundreds kilometers large) with a very small mesh size (of
order few tens of meters). This would make computations and data storing
practically untractable. This is why we have defined a procedure which
computes data only on very small subgrids whose size (and mesh size)
depends on the wavelet scale and on the event size. This procedure is made
possible because we compute wavelet coefficients at several scales but only
at a single location, namely the position of the upcoming large event.
Indeed, we do not store the stress history for all locations, but only at the position
$P_0$ of the epicenter of the target main shock.

\section{Results}

We have analyzed the evolution of the stress field before $4$ large Southern
Californian shocks: Loma Prieta (1989), Landers (1992), Northridge
(1994) and Hector Mine (1999). We restricted our analysis to those $4$
recent events as this ensures that our computed stress field history is
the longest possible for this area, and is not subjected to finite size
effects (as these 4 events are located at the end of the catalogue). The Caltech
catalog we used is thought to be complete since 1932 for events of
magnitude larger than $3.5$. Computation of the stress field before each
of the selected large events included all events of magnitude larger
than $4$ since 1932. 

Three parameters dictate the properties of the Green function of a
seismic event in our computations, namely the relaxation time scales
$\tau_1$ and $\tau_2$, and the stress amplification factor $B$. We made
several computations, varying those $3$ parameters. We checked that the
less influential parameter is $B$. Another parameter which has a rather
low influence on the results is $\tau_2$, the relaxation time of the
lower, more ductile medium. The most influential parameter is $\tau_1$,
the relaxation time of the upper layer. If $\tau_1$ is too small, then
all events appear as very well individualized temporal stress pulses
decaying very fast before the next event takes place. As a consequence,
the dominating space scale is never defined, except at the time of
occurrence of each event, where it is of the order of the distance
between this event and $P_0$. The optimal space scale thus varies very
wildly with time. 

When increasing $\tau_1$, stress pulses gradually
overlap in time. Finally, when $\tau_1$ is infinite, stress pulses becomes
steps without any relaxation. Increasing $\tau_1$ leads to a less
erratic behaviour of the optimal spatial scale length obtained from
our wavelet analysis. We will here
consider a Green function with relaxation times $\tau_1 =100$
years and $\tau_2 =0.5$ year. The scalar stress history computed at the
location of the Landers shock is shown on Figure $2c$. It globally
increases with time (as all previous events stress perturbations are
positive by definition) but does not exhibit any acceleration. Note that
stress steps (due to neighboring events) are followed by a smooth
decay, due to the very slow relaxation associated with the high $\tau_1$ value. 
The time step for the computation of each successive
point of the cumulative stress is 6 months. We stress that
the procedure we use provides results independent of the time
step, thanks to our linear rheology. 

Figure $3$ shows the wavelet
coefficients for the cumulative stress function
constructed for the Landers 1992 earthquake
as a function of scale at various times. Time increases
from the bottom to the top (the very upper
curve has been computed just before the Landers shock). 
The curves with
the lowest amplitudes, corresponding to
the early years, are flat as the number of shocks is low, so that the
stress field is almost $0$ everywhere, and no specific structure emerges
as too few events have been included in the computation. Later, the
amplitude of the
profile increases in amplitude (either positively or negatively), but it
is worth noting that its shape is almost constant. As time increases,
the amplitude of the stress field varies, but its structure remains
constant,
at least at point $P_0$. For example, for wavelet scales lower than
$10km$ (true size lower than $22km$),
the ``future'' Landers epicenter is found to be located in a local stress deficit.
The local correlation length of the stress field, given by the maximum
of the wavelet coefficient, occurs for a constant scale of about $25km$ (true size of
about 55 km). We note that this maximum occurs at the same scale for all times.
Figure $4$ shows the evolution with time of the correlation length. It first
fluctuates widely, as there are too few events to
compute a representative stress fluctuations field, but then enters a
very stable phase with no noticeable variation with time. We thus show
no increase or decrease of this local correlation length, which confirms
the fact that the
local structure of the computed stress field does not exhibit any major
change when approaching failure around $P_0$.

Figures $5$ to $7$ show the results of the same computations before the
Loma Prieta event. The correlation length
is found constant from $1958$ to $1987$, with a value of about $77 km$
(wavelet scale of $35 km$).

Figures $8$ to $10$ show the results of the same computations before the
Northridge event. The correlation length is found
constant from 1972 to 1994, with a value of about
$66 km$. 

Figures $11$ to $13$ show the results of the same computations before the
Hector Mine 1999 event. Once again, no clear increase of the
correlation length occurs before the large event. 

We also performed the
same tests considering only
catalog events of magnitude larger than $5$. We obtain exactly the same
results, except that the wavelet profiles
of Figures $3, 6, 9$ and $12$ are found to be dilated along the scale axis. This
just reflects the fact fewer events are taken
into account in the computations, and are thus more diluted in space. We
also performed tests using a
larger distance of influence of each given event 
quantified by equation (\ref{mgksl})) by doubling the
rupture size $L \to 2L$. The results are qualitatively the same.

\section{Interpretation and discussion}

Using simplified models of earthquake elastic stress transfer and of
the lithosphere rheology, we have attempted to model the stress field evolution
from 1932 up to the occurrrence time of recent large Southern
Californian events. This allowed us to analyze the time evolution of our
simplified cumulative stress field at the locus of large impending shocks before
their occurrence, and to determine the spatial correlation length
of this local stress field. Using a variety of rheological models did not
allow us to find evidence of a strong increase (nor any other peculiar
variation) of both the cumulative stress
field and of the correlation length before any of the $4$ major events
studied here. These negative results would not change by replacing 
the simple exponential decays by power laws of the form of the Omori law
for aftershocks, since taking an infinite range correlation 
$\tau_1 \to +\infty$ does not change our results.

We have observed that all large events occured in a local minimum of the
computed stress field at (true) scales less than $20-25km$, and that
this minimum becomes more and more pronounced with time. A magnitude $7$
event has an average rupture length of about $70km$. As we have stressed
before, such an event certainly nucleates in a zone where the stress
field is correlated on long wavelengths. The final length of the rupture
will stem from the interplay between this initial static stress field
structure and details of rupture dynamics (inertial effects coupled with
the specific geometry of the rupture plane). We can guess that the final
extent of the rupture will be larger than the initial correlation length
of the stress field. This is why we could expect that this correlation
length before each of the $4$ major events should have been of the order of a few
tens of kilometers. It is thus puzzling to observe that the wavelet
coefficients at scales of $10$ to $20 km$ are becoming more and more
negative with time. This observation is perhaps due to the naive shape
of the Green function we considered, which is positive everywhere.
However, we believe that if this assumption certainly affects the value
of the computed stress field, it should certainly lead us to an
overestimation of the correlation length, as more space is filled with
positive stress. We are thus forced to conclude that there is neither
a strong stress field and nor large stress
correlation at the scale of a few kilometers scale. It thus
seems that the mechanism of stress transfer due to the
occurrence of successive smaller-sized events is not a direct ingredient
in building long correlations in the cumulative stress field, which are
necessary for the propagation of large future events according to the
critical point model.

These results are in contradiction with those reported
in the literature [Bufe and Varnes, 1993; Sornette and Sammis, 1995; Bowman et al., 1998;
Brehm and Braile, 1998; 1999; Jaum\'e and Sykes, 1999; Ouillon and Sornette, 2000] based
on the cumulative Benioff strain, 
who showed that large-scale spatial and temporal correlations
characterize seismicity before a large event in the same area. 

Our results may be reconciled with those previous 
studies if we acknowledge that medium-sized events
are not seismo active (they are not ``actors''). In other words,
the temporal singularities
defined in [Bowman et al., 1998] for instance stem rather from the large scale
geometry of the boundary loading conditions and correlations not directly
mediated by the stress field (that were not taken into
account in the present work) than from strong interaction between
seismic events mediated by the stress field. In this spirit,
King and Bowman [2001] and Sammis and Sornette [2002]
have developed a model in which the main mode of loading of a
previously ruptured major fault occurs by localized viscous flow
beneath this fault. The consequence is that the extent of the stress
shadow due to the previous mainshock decreases with time, so that
seismicity migrates back to the mainshock epicenter in an accelerating
manner, the temporal singularity coinciding with a new mainshock on the
fault. However, such a model implies that seismicity migrates towards $P_0$,
which cannot reasonably be inferred from our computations either (Figure
$3, 6, 9$ and $12$). If this was the case, the wavelet coefficients should be
negative at $P_0$, and the width of the domain around $P_0$ where
coefficients are negative should decrease with time. This suggests
that the loading mechanism proposed by King and Bowman [2001] and Sammis and Sornette [2002]
does not explain the data, but that another loading mechanism may explain the
temporal singularity coinciding with large events.

Another solution to explain the discrepancy between the large scale
correlations observed in seismic catalogs 
[Bufe and Varnes, 1993; Sornette and Sammis, 1995; Bowman et al., 1998;
Brehm and Braile, 1998; 1999; Jaum\'e and Sykes, 1999; Ouillon and Sornette, 2000]
and our results is to argue that our geometrical/rheological model of the lithosphere
is incorrect, which makes our Green function imperfect.
The Green function we have
considered is representative of a linear viscoelatic layered medium, and
we checked that our results are not strongly dependent on its
various parameters. One possibility is that, if the observed absence
of correlations is due to our choice of the Green function, then
the true Green function must be of a fundamentally different nature. The
Earth's crust is a very complex medium, composed of blocks of various
sizes separated by fractures or fault zones, subjected to a confining
pressure and temperature increasing with depth. We would be indeed very
lucky if such a medium behaved as a perfect linear medium. Indeed,
crustal rheology must be of nonlinear nature, even in its most
superficial ``elastic'' part. Some evidence of a nonlinear response 
associated with the anisotropic response of a cracked medium under compression
compared to tension has been reported in [Peltzer et al., 1999].
Extending this argument, 
if, for example, the crust behaves as a
granular material, then we must expect that tectonic forces propagate
over longer distances within much narrower channels than those predicted
by standard elastic models. This singular property is due to the
hyperbolic nature of stress propagation differential equations in
granular media [Bouchaud et al., 1995; 2001], 
whereas those equations are of ellipitical nature in
standard elasto-plastic media. The 
real rheology of the Earth's crust is probably
somewhere between that of a granular material and standard
(possibly nonlinear) visco-elastico-plasticity. It thus seems 
important to better understand crustal rheology (and its associated
Green function), in order to check the changes it would imply in the
various brittle crustal modes of deformation and in the way earthquakes
``speak to each other.'' In this spirit, phenomenological models of
earthquake interaction and triggering are quite successful in capturing
most of the phenomology of seismic catalogs
[Helmstetter and Sornette, 2002; Helmstetter et al., 2002]. It remains to derive
the triggering Green function from physically-based mechanisms, which 
seem to require much more than just visco-elastic stress transfers.
     
We thank A. Helmstetter for useful discussions and for a critical reading
of the manuscript.

\section*{References}

\vskip 0.3cm
\noindent All\`egre, C.J., Le Mouel, J.L. and Provost, A., Scaling rules
in rock fracture  and possible implications for earthquake predictions,
Nature  297, 47-49, 1982.

\vskip 0.3cm
\noindent Bouchaud, J.P., Cates, M.E. and Claudin, P.,
Stress distribution in granular media and nonlinear wave equation,
Journal de Physique I 5, 639-656, 1995.

\vskip 0.3cm
\noindent Bouchaud, J.P., Claudin, P., Levine, D. and Otto, M.,
Force chain splitting in granular materials: A mechanism for large-scale
pseudo-elastic behaviour, European Physical Journal 4, 451-457, 2001.

\vskip 0.3cm
\noindent Bowman, D.D. and King, G.C.P.,
Accelerating seismicity and stress accumulation before large earthquakes,
Geophys. Res. Lett. 28, 4039-4042, 2001.

\vskip 0.3cm
\noindent Bowman, D.D. and King, G.C.P.,
Stress transfer and seismicity changes before large earthquakes,
Comptes Rend. Acad. Sci. II (A) Paris, 333, 591-599, 2001.

\vskip 0.3cm
\noindent Bowman, D. D., Ouillon, G., Sammis, C. G., Sornette, A., and Sornette,
D., An observational test of the critical earthquake con-cept,
J. Geophys. Res., 103, 24 359Ð24 372, 1998. 

\vskip 0.3cm
\noindent Brehm, D.J. and Braile, L.W.,
Intermediate-term earthquake prediction using precursory events in the new
Madrid seismic zone, Bull. Seism. Soc. Am. 88, 564-580, 1998.

\vskip 0.3cm
\noindent Brehm, D.J. and Braile, L.W.,
Intermediate-term earthquake prediction using the modified time-to-failure
method in southern California, Bull. Seism. Soc. Am. 89, 275-293, 1999.

\vskip 0.3cm
\noindent Bufe, C. G. and Varnes, D. J.: Predictive modeling of the seismic
cycle of the greater San Francisco Bay region, J. Geophys. Res., 98,
9871Ð9883, 1993.

\vskip 0.3cm
\noindent Gross, S. and Rundle, J.,
A systematic test of time-to-failure analysis, Geophys. J. Int. 133, 57-64, 1998.

\vskip 0.3cm
\noindent Guarino, A., Ciliberto, S., Garcimartin, A., Zei, M. and R. Scorretti,
Failure time and critical behaviour of fracture precursors in
heterogeneous materials, Eur. Phys. J. B 26, 141-151, 2002.

\vskip 0.3cm
\noindent Helmstetter, A. and D. Sornette,
Sub-critical and super-critical regimes in epidemic models of
earthquake aftershocks, in press in to J. Geophys. Res., 2002
(http://arXiv.org/abs/cond-mat/0109318)

\vskip 0.3cm
\noindent Helmstetter, A. and D. Sornette, Diffusion of earthquake 
aftershock epicenters and Omori's law: exact mapping to generalized 
continuous-time random walk models, in press to Phys. Rev. E, 2002  
(http://arXiv.org/abs/cond-mat/0203505)

\vskip 0.3cm
\noindent Helmstetter, A., D. Sornette and J.-R. Grasso,
Mainshocks are Aftershocks of Conditional Foreshocks: How do foreshock
statistical properties emerge from aftershock laws, 2002
submitted to J. Geophys. Res.  (http://arXiv.org/abs/cond-mat/0205499)

\vskip 0.3cm
\noindent Huang, Y., H. Saleur, C. G. Sammis, D. Sornette,
Precursors, aftershocks, criticality and self-organized criticality, 
Europhysics Letters 41, 43-48, 1998.

\vskip 0.3cm
\noindent Jaum\'e, S.C. and L.R. Sykes, Evolving toward a critical point: 
 a review of accelerating seismic moment/energy release 
 prior to large and great earthquakes, Pure appl. geophys., 155, 279-306, 1999.

\vskip 0.3cm
\noindent Johansen,~A.,~Sornette,~D.,~Wakita,~G.,~Tsunogai,~U.,~Newman,~W.I. and
Saleur,~H., Discrete scaling in earthquake precursory phenomena: evidence in the Kobe
earthquake,~Japan, J. Phys. I France, 6,~1391-1402, 1996.

\vskip 0.3cm
\noindent Johansen, A., H. Saleur and D. Sornette,
New Evidence of Earthquake Precursory Phenomena in the
17 Jan. 1995 Kobe Earthquake, Japan, Eur. Phys. J. B 15, 551-555, 2000.

\vskip 0.3cm
\noindent Johansen, A. and D. Sornette, Critical Ruptures, Eur. Phys. J., B 18,
163-181, 2000.

\vskip 0.3cm
\noindent Jones, L. M. and P. Molnar, Some characteristics of foreshocks and
	their possible relationship to earthquake prediction and premonitory
	slip on fault, J. Geophys. Res. 84, 3596-3608, 1979.
	
\vskip 0.3cm
\noindent Keilis-Borok, V.I., The lithosphere of the earth as a nonlinear system
with implications for earthquake prediction, Reviews of Geophysics 28 N1, 19-34, 1990.

\vskip 0.3cm
\noindent Keilis-Borok, V.I. and L.N. Malinovskaya, One regularity in the occurrence
of strong earthquakes, J. Geophys. Res. 69, 3019-3024, 1964.

\vskip 0.3cm
\noindent Keilis-Borok, V.I. and A.A. Soloviev, Nonlinear dynamics of the 
lithosphere and earthquake prediction (Springer-Verlag, Heidelberg, 2002).

\vskip 0.3cm
\noindent Knopoff, L., Levshina, T., Keilis-Borok, V. and
Mattoni, C., Increased long-range intermediate-magnitude earthquake 
activity prior to strong earthquakes in California,
J. Geophys. Res. 101, 5779-5796, 1996.

\vskip 0.3cm
\noindent Mora, P. et al., in ``Geocomplexity and the Physics of
Earthquakes'', eds Rundle, J. B., Turcotte, D. L. \& Klein, W. (Am. Geophys. Union,
Washington, 2000).

\vskip 0.3cm
\noindent Mora, P. and D. Place (2002) 
Microscopic simulation of stress correlation
evolution: implication for the critical point hypothesis for
earthquakes, Pure Appl. Geophys., in press (preprint at
http://www.quakes.uq.edu.au/research/PDF/Mora1PAG01.pdf).

\vskip 0.3cm
\noindent Newman, W.I., D.L. Turcotte and A.M. Gabrielov,
Log-periodic behavior of a hierarchical failure model with applications to
precursory seismic activation, Phys.Rev. E 52, 4827-4835, 1995.

\vskip 0.3cm
\noindent Ouillon, G., Application de l'Analyse Multifractale et de la
Transform\'ee en OndelettesAnisotropes à la Caractérisation Géométrique
Multi-Echelle des Réseaux de Failles et deFractures, Editions du BRGM,
1995.

\vskip 0.3cm
\noindent Ouillon, G. and D. Sornette,
The critical earthquake concept applied to mine rockbursts with time-to-failure analysis, 
Geophysical Journal International 143, 454-468, 2000.

\vskip 0.3cm
\noindent Peltzer, G., Crampe, F. and King, G.,
Evidence of nonlinear elasticity of the crust from the Mw7.6 Manyi (Tibet) earthquake,
Science 286 N5438, 272-276, 1999.

\vskip 0.3cm
\noindent Reynolds, P.J., Klein, W. and Stanley, H.E.,
Renormalization Group for Site and Bond Percolation, J. Phys. C 10, L167-L172, 1977.

\vskip 0.3cm
\noindent Rundle, J.B., Jackson D.D., A Three-Dimensional Viscoelastic Model of a
Strike Slip Fault, Geophys. J. R. ast. Soc., 49, 575-591, 1977.

\vskip 0.3cm
\noindent Saleur,~H., Sammis,~C.G. and Sornette,~D.,
Renormalization group theory of earthquakes, 
Nonlinear Processes in Geophysics 3,~102-109, 1996.

\vskip 0.3cm
\noindent Saleur,~H.,~Sammis,~C.G. and Sornette,~D.,
Discrete scale invariance,~complex fractal dimensions and log-periodic
corrections in earthquakes, J. Geophys. Res., 101,~17661--17677, 1996.
 
\vskip 0.3cm
\noindent Smalley, R.F., D.L. Turcotte, and S.A. Sola, A renormalization group
approach to the stick-slip behavior of faults, J. Geophys. Res. 90,
1884-1900, 1985.

\vskip 0.3cm
\noindent Sobolev, G.A., Y.S. Tyupkin, Analysis of Energy Release Process during
Main Rupture Formation in Laboratory Studies of Rock Fracture and before
Strong Earthquakes, Phys. Solid Earth, 36, 2, 138-149, 2000.

\vskip 0.3cm
\noindent Sornette, D.,
Elasticity and failure of a set of elements loaded 
in parallel, J.Phys.A, 22, L243-250, 1989a.

\vskip 0.3cm
\noindent Sornette, D., 
Acoustic waves in random media: II Coherent effects and strong disorder regime,  Acustica  67,  
251-265, 1989b.

\vskip 0.3cm
\noindent Sornette, D.,
Acoustic waves in random media: III Experimental situations,  Acustica 68, 15-25, 1989c.

\vskip 0.3cm
\noindent Sornette, D., Critical Phenomena in Natural Sciences, 
Chaos, Fractals, Self-organization and Disorder: Concepts and Tools,
(Springer Series in Synergetics, Heidelberg, 2000).

\vskip 0.3cm
\noindent Sornette, D. and V.F. Pisarenko,
Fractal Plate Tectonics, in press Geophysical Research Letters, 2002.
(preprint at http://arXiv.org/abs/cond-mat/0202320)

\vskip 0.3cm
\noindent Sornette, D. and C.G. Sammis, Complex critical exponents from
renormalization group theory of earthquakes : Implications for
earthquake predictions, J.Phys.I. France, 5, 607-619, 1995.

\vskip 0.3cm
\noindent Sornette,~A. and Sornette,~D., Earthquake rupture as a
critical point: Consequences for telluric precursors, Tectonophysics
179, 327-334, 1990.

\vskip 0.3cm
\noindent Souillard, B., in 
Chance and matter, edited by J. Soulettie, J.
Vannimenus and R. Stora (Amsterdam; New York, N.Y., Elsevier Pub. Co., North-Holland, 1987).

\vskip 0.3cm
\noindent Sykes L.R. and S. Jaum\'e, Seismic activity on neighboring faults as a
long-term precursor to large earthquakes in the San Francisco Bay Area,
Nature, 348, 595-599, 1990. 

\vskip 0.3cm
\noindent Vere-Jones, D., Statistical theories of crack
propagation, Mathematical Geology 9, 455-481, 1977.

\vskip 0.3cm
\noindent Voight,~B., A method for prediction of volcanic eruptions,
Nature 332, 125-130, 1988.

\vskip 0.3cm
\noindent Voight,~B., A relation to describe rate-dependent material
failure, Science 243, 200-203, 1989.

\vskip 0.3cm
\noindent Wells, D.L., K.J. Coppersmith, New Empirical Relationships among
Magnitude, Rupture Length, Rupture Width, Rupture Area, and surface
Displacement, Bull. Seism. Soc. Am, 84, 974-1002, 1994.

\vskip 0.3cm
\noindent Zoller. G., and S. Hainzl,
Detecting premonitory seismicity patterns based on critical point
dynamics, Natural Hazards and Earth System Sciences, 1, 93-98, 2001.

\vskip 0.3cm
\noindent Zoller, G. and S. Hainzl,
A systematic spatiotemporal test of the critical point hypothesis for
large earthquakes, Geophys. Res. Lett., 29 (11), 10.1029/2002GL014856, 2002.

\vskip 0.3cm
\noindent Zoller, G., S. Hainzl, and J. Kurths,
Observation of growing correlation length as an indicator for critical
point behavior prior to large earthquakes, J. Geophys. Res., 106, 2167-2176, 2001.
%http://www.agnld.uni-potsdam.de/~hainzl/publications.html

\pagebreak

\vskip 0.3cm
Figure 1: Evolution with time of the time-dependent 
part of the normalized stress field showing the loading phase induced by the relaxing
lower layer and the large time relaxation phase in the upper layer.
The parameters are $\tau_1 =10$ years, $\tau_2 =1$ year and $B=1$.

\vskip 0.3cm
Figure 2: Cumulative stress function as a function of time
at the location of the Landers epicenter 
calculated by summing the contributions
$\sigma(r_i,t_i)$ given by (\ref{mgjslala}) of the Green functions generated
by all previous events $i$, that occurred at times $t_i$ prior to the Landers earthquake
taken at the origin of time and at distances $r_i$ from the Landers epicenter.
(a) $\tau_1 =1$ year and $\tau_2 =6$ months; (b) $\tau_1 = 10$ years
and $\tau_2 =6$ months; (c) Same as Figure 2a with  
$\tau_1 = 100$ years and $\tau_2 =6$ months.

\vskip 0.3cm
Figure 3: Wavelet
coefficients for the cumulative stress function
constructed for the Landers 1992 earthquake
as a function of scale $a$ at various times. Time increases
from the bottom to the top (the very upper
curve has been computed just before the Landers shock). 

\vskip 0.3cm
Figure 4: Correlation length estimated at the Landers epicenter
of the cumulative stress function for the Landers
earthquake as a function of time.

\vskip 0.3cm
Figure 5: Same as Figure 2c for the Loma Prieta 1989 earthquake.

\vskip 0.3cm
Figure 6: Same as Figure 3 for the Loma Prieta 1989 earthquake.

\vskip 0.3cm
Figure 7: Same as Figure 4 for the Loma Prieta 1989 earthquake.
The correlation length
is found constant from $1958$ to $1987$, with a value of about $77 km$
(wavelet scale of $35 km$).

\vskip 0.3cm
Figure 8: Same as Figure 2c for the Northridge 1994 earthquake.

\vskip 0.3cm
Figure 9: Same as Figure 3 for the Northridge 1994 earthquake.

\vskip 0.3cm
Figure 10: Same as Figure 4 for the Northridge 1994 earthquake.
The correlation length is found
constant from 1972 to 1994, with a value of about
$66 km$. 

\vskip 0.3cm
Figure 11: Same as Figure 2c for the Hector Mine 1999 earthquake.

\vskip 0.3cm
Figure 12: Same as Figure 3 for the Hector Mine 1999 earthquake.

\vskip 0.3cm
Figure 13: Same as Figure 4 for the Hector Mine 1999 earthquake.

\end{document}